\documentclass[prd,preprint,showpacs]{revtex4}
\usepackage{amsmath}
\usepackage{amssymb}

\begin{document}

\title{A theory of evolving natural constants embracing Einstein's
theory of general relativity and Dirac's large number hypothesis}

\author{H. W. Peng}

\email[Email address: ] {penghw@itp.ac.cn} \affiliation{Institute
of Theoretical Physics, Chinese Academy of Sciences,
 P.O.Box 2735 Beijing 100080, China}

\begin{abstract} Taking a hint from Dirac's large number hypothesis, we note the
existence of cosmic combined conservation laws that work to
cosmologically long time. We thus modify or generalize Einstein's
theory of general relativity with fixed gravitation constant $G$
to a theory for varying $G$, which can be applied to cosmology
without inconsistency, where a tensor arising from the variation
of G takes the place of the cosmological constant term. We then
develop on this basis a systematic theory of evolving natural
constants $m_{e},m_{p},e,\hslash ,k_{B}$ by finding out their
cosmic combined counterparts involving factors of appropriate
powers of $G$ that remain truly constant to cosmologically long
time. As $G$ varies so little in recent centuries, so we take
these natural constants to be constant.
\end{abstract}
 \maketitle

\section{General relativity and large number hypothesis}

In Einstein's theory of general relativity as in Newton's theory of
gravitation, the strength of the gravitational interaction is described by a
fixed dimensional constant, namely$\ $the Newtonian gravitation constant $\
G_{N}\doteq 6.7\times 10^{-11}$m$^{\text{3}}$kg$^{-1}$s$^{-\text{2}}$.
Together with the velocity of light in vacuo $c\doteq 3\times 10^{8}$m/s
which serves to convert time into length by putting $c=1$, there occur also
in Einstein's theory $G_{N}/c^{2}\doteq 7.4\times 10^{-28}$mkg$^{-1}$ and $%
G_{N}/c^{4}\doteq 8.2\times 10^{-45}$mJ$^{-1}$ which respectively serve to
convert mass and energy into length by putting also $G_{N}=1$. Einstein's
theory has been applied to cosmology; in addition to the dimensional
constants $G_{N}$\ another cosmological constant with dimension of length to
the power minus two, was once specially introduced by Einstein in his
attempt to construct a static model for the universe. This soon has lost his
favor as observations of the red shifts of extra-galactic nebulae show
defintely that the universe is expanding, but as an additional parameter,
cosmological constant is still used today in trying to fit the observational
data with the non-static homogeneous cosmological model using the
Roberston-Walker metric. Rough estimates have thus been obtained for the age
$t$\ of universe at present to be $\approx 10^{9}$ years or $\approx 10^{10}$%
\ years more recently.\

On comparing the ratio $e^{2}/(Gm_{p}m_{e})$ of the electrostatic to the
gravitational force between the proton and the electron in an hydrogen atom,
which is a large dimensionless number of the order of $10^{39}$ , with the
ratio $t/(e^{2}/m_{e}c^{3})$ of the age $t$\ of universe at present to the
time needed for light to travel a distance of the classical radius of the
electron, which is also a large dimensionless number of the order of $%
10^{39} $ or$\ 10^{40}$, one proposes \cite{Dirac1} that a large
number equation

\begin{equation*}
\bigskip e^{2}/(Gm_{p}m_{e})\approx t/(e^{2}/m_{e}c^{3})\text{ \ \ \ \ (1.1)}
\end{equation*}%
should hold. Further from Hubble's estimate $\rho =(1.3$ to $1.6)\times
10^{-30}$ g/cm$^{3}$ for the density of matter due to the extra-galactic
nebulae averaged over cosmic space and a factor thousand or hundred times to
include dust or invisible matter, one can estimate the ratio $M/m_{p}$ of
the total mass $M$ of matter in the universe at present of radius $R$ to the
proton mass, which is roughly of the order $10^{78}$ or $10^{80}$, one
proposes \cite{Dirac1} that another large number equation%
\begin{equation*}
\ M/m_{p}\approx [t/(e^{2}/m_{e}c^{3})]^{2}\text{ \ \ \ (1.2)}
\end{equation*}%
\ should also hold. By a large number equation we mean wherein some unknown
factors of small numbers\ are understood to be there but not written out.
Eddington takes seriously the relation%
\begin{equation*}
M/m_{H}=N\approx \bigskip \lbrack e^{2}/(Gm_{p}m_{e})]^{2}\approx 10^{78}%
\text{ \ (1.3)}
\end{equation*}%
and advocates in his book ``Fundamental Theory" that in the
uranoid (model of universe with $N$\ hydrogen atoms) when one
studies one particular hydrogen atom the other $N-1$ hydrogen
atoms must also be considered as representing the rest of the
universe.

Based on these empirical large number equations (1.1) and (1.2)
Dirac proposes the large number hypothesis \cite{Dirac1} that
during the evolution of the universe the strength $G$\ of
gravitational interaction gradually weakens as time $t$ goes on
according to$\ G\varpropto t^{-1}$\ while mass
is continuously being created and so increasing with time according to $%
M\varpropto t^{2}$ . He reached at this hypothesis by adding an additional
assumption that the atomic constants remain unchanged and do not evolve. One
can, however, arrive at Dirac's large number hypothesis by the following
evolving atomic constants, say, both $m_{p}$ and $m_{e}\varpropto t^{2}$\
while $e^{2}\varpropto t^{3}$\ , leaving $c$\ alone not evolving. This
corresponds to Eddington's assumption that $N$ remains unchanged.

In this paper we shall take the large number equations, though empirical at
the present stage of knowledge, rather seriously. On dividing (1.2) by (1.1)
we obtain a relation free from all atomic constants $e,m_{e},m_{p},$ \
\begin{equation*}
GM\approx c^{3}t\text{ \ \ \ (1.4)}
\end{equation*}%
which exhibits clearly the inconsistency of applying Einstein's
theory of general relativity to cosmology. If one regard this
result, though empirical, as true, then clearly $G$\ and $M$\
cannot both conserve in cosmologically long time, and this is in
contradiction to the assumptions inherent in Einstein's theory of
general relativity that $G$\ be constant and matter do conserve.
Hence one needs to modify Einstein's theory for application to
cosmology. But we know, for phenomena including the crucial tests
that occur, from the cosmic point of view, in a small neighborhood
and for short duration, Einstein's theory is a good approximation.
This has been confirmed by observations and experiments and must
not be spoiled. We
note, according to Dirac's large number hypothesis, it is neither $M\ $nor $%
G $\ but$\ $the combination\ $G^{2}M$ that conserves in cosmologically long
time. We shall take such cosmic combined (c.c.) conservation law working to
cosmologically long time as a hint in our attempts to modify or rather to
generalize Einstein's theory of general relativity, to be consistent with
Dirac's large number hypothesis, and develop systematically a theory of
evolving natural constants. In so doing we find that it is possible to
generalize Dirac's large number hypothesis somewhat to $G\varpropto t^{-n}$\
with $n$ not necessarily equal to one. It seems that c. c. conservation laws
can be more simply found and stated, but the variation of $G$,\ together
with that of $g_{\mu \nu }$, as functions of space-time must be left to the
solution of various peoblems.

\section{Theory with varying $G$}

The problem of non-static homogeneous matter-dominated cosmological model is
a simple case to work with mathematically. Because there is only one
independent variable $t$, it is easier here to find out how natural
constants shall evolve and how the usual conservation laws shall be combined
to work in cosmologically long time. We shall not make Dirac's assumption
that atomic constants do not change in the long. (In this paper by long or
short we mean cosmologically long or short time and similarly by large or
small we mean cosmologically large or small distance). But we do follow the
spirit of Dirac's large number hypothesis and generalize it into
\begin{equation*}
\ G\varpropto t^{-n},(0<n<2),\text{ so }t\varpropto G^{-1/n}\ \ \text{(2.1)}
\end{equation*}%
where $n$\ can be, but not necessarily, equal to one. Then (1.3) leads to
\begin{equation*}
G^{1+1/n}M\varpropto t^{0}\text{ i.e. }G^{1+1/n}M=\widetilde{M}\text{\
conserves.\ (2.2)}
\end{equation*}%
To save one's worry about the dimension of the cosmic combined (c.c.)
quantity like $\widetilde{M}$, it would be most simple to introduce a
constant but arbitrary dimensional conversion factor $G_{0}$\ to convert
mass into length by putting it to be $G_{0}=1$, so that in (2.2) and similar
equations in the following we understand that the c. c. quantity (like $%
\widetilde{M}\ $) is always of the same dimension as the original quantity
(like $M$). Thus the equation (2.2) is meant to be, when writen in full,
\begin{equation*}
(G/G_{0})^{1+1/n}M=\widetilde{M}\text{ \ (2.2a)}
\end{equation*}%
As usual we take $G_{N}$ for $G_{0}$.\ In order to conform to such
a c. c. conservation law for $\widetilde{M}\ $\ we modify the
corresponding action integral for simple matter without internal
stress which in Einstein's theory is given, as shown in Dirac's
book \cite{Dirac2} on general theory of relativity, by
\begin{equation*}
I_{m}=-\int (g_{\mu \nu }\text{$\wp $}^{\mu }\text{$\wp $}^{\nu
})^{1/2}d^{4}x\text{ \ (E2.3a) with constraint $\wp $}^{\mu },_{\mu }=0\text{
\ (E2.3b)}
\end{equation*}%
into a similar expression%
\begin{equation*}
\widetilde{I}_{m}=-\int (g_{\mu \nu }\widetilde{\text{$\wp $}}^{\mu }%
\widetilde{\text{$\wp $}}^{\nu })^{1/2}d^{4}x\text{ (2.3a) with modified
constraint }\widetilde{\text{$\wp $}}^{\mu },_{\mu }=0\text{ \ (2.3b)}
\end{equation*}%
Here with a change of notation
\begin{eqnarray*}
u^{\mu } &=&\text{$\wp $}^{\mu }/(g_{\alpha \beta }\text{$\wp $}^{\alpha }%
\text{$\wp $}^{\beta })^{1/2}\text{ \ (E2.4a), so }u^{\mu }u_{\mu }=1 \\
\text{\ }(g_{\alpha \beta }\text{$\wp $}^{\alpha }\text{$\wp $}^{\beta
})^{1/2} &=&\rho _{m}\surd (-g)\text{ \ so $\wp $}^{\mu }=u^{\mu }\rho
_{m}\surd (-g)\text{ \ (E2.4b)}
\end{eqnarray*}%
the constraint in Einstein's theory gives the conservation of mass and may
be written as the equation of continuity
\begin{equation*}
\int \text{$\wp $}^{4}dx^{1}dx^{2}dx^{3}=M\text{ \ (E2.4c) }(\rho _{m}u^{\mu
})_{;\mu }=0\text{ (E2.4d)}
\end{equation*}%
Similarly with a change of notation with the cosmic combined quantities
\begin{equation*}
(g_{\alpha \beta }\widetilde{\text{$\wp $}}^{\alpha }\widetilde{\text{$\wp $}%
}^{\beta })^{1/2}=\widetilde{\rho }_{m}\surd (-g)\text{ \ (2.4a) \ }%
\widetilde{\text{$\wp $}}^{\mu }=u^{\mu }\widetilde{\rho }_{m}\surd (-g)%
\text{ \ (2.4b) }
\end{equation*}%
where
\begin{equation*}
\text{\ }\widetilde{\rho }_{m}=G^{1+1/n}\rho _{m}\text{ \ (2.5)}
\end{equation*}%
the modified constraint (2.3b) in the modified theory gives the conservation
of the cosmic combined mass
\begin{equation*}
\int \widetilde{\text{$\wp $}}^{4}dx^{1}dx^{2}dx^{3}=\widetilde{M}=G^{1+1/n}M%
\text{ \ \ (2.4c)(2.5c)}
\end{equation*}%
for homogeneous $G=G(t)$ and the corresponding equation of continuity
\begin{equation*}
(\widetilde{\rho }_{m}u^{\mu })_{;\mu }=0\text{ (2.4d)}
\end{equation*}%
\ In case of a mass point described with a three dimensional delta function
factor in (E2.4c) and (2.4c), we obtain from (2.5c) the cosmic combined
constant mass%
\begin{equation*}
\widetilde{m}=G^{1+1/n}m\text{ \ (2.5a)}
\end{equation*}%
which holds for a proton, an electron, an atom or a molecue, contrary to
Dirac's assumtion that atomic constants like $m_{e}$\ and $m_{p}$\ do not
evolve. From this we see that our modification or generalization of
Einstein's theory consists of multiplying the integrand of the source action
integral by the dimensionless varying factor $G^{1+1/n}$ and introducing new
variables suitable for expressing the modified cosmic combined conservation
law. This rule will be kept throughout for other action integrals involved
in the comprehensive action principle, cf. \S 6, \S 8 and \S 12 later.

As to Einstein's original gravitational action integral where $G^{-1}$ is
constant
\begin{equation*}
I_{g}=(16\pi )^{-1}\int G^{-1}R_{\sigma }^{\sigma }\surd (-g)d^{4}x\text{
where }R_{\upsilon }^{\sigma }=R_{\mu \nu }g^{\mu \nu }\text{ \ \ (E2.6)}
\end{equation*}%
\ we multiply the integrand by the factor $G^{1+1/n}$ and add\ a kinetic
term thus:%
\begin{equation*}
\widetilde{I}_{g}=(16\pi )^{-1}\int [G^{1/n}R_{\sigma }^{\sigma
}-w(G^{1/(2n)})_{;\sigma }(G^{1/(2n)})^{;\sigma }]\surd (-g)d^{4}x\text{ \ \
(2.6)}
\end{equation*}%
This is the most general expression for an action integral involving only
field variables $G,g_{\mu \nu }$\ without any dimensional constants that
will reduce to Einstein's $I_{g}$ in case $G$\ be a constant. As the
numerical factor $(16\pi )^{-1}$ in Einstein's theory is determined by
comparison with Newton's theory in the so called non-relativistic
approximation (which is good for phenomena with velocity small compared with
that of light), so the factor $(16\pi )^{-1}$ in our modified theory is
determined by comparison with Einstein's theory in the non-cosmologic
approximation (which is good for phenomena of duration and distance short
compared with that of the universe, as will be shown in \S 5). The numerical
constant $w$\ will be determined from the non-static homogeneous
matter-dominated cosmological model\ in \S 4 by using our generalized large
number hypothesis (2.1). We prefer the choice of $w=8$\ for the case $k=0$
of the Robertson-Walker metric.

In later sections we shall see that such a modification or generalization
can be carried through other source action integrals like that for Maxwell's
theory of electromagnetism (\S 6) or that for Dirac's theory of electron (\S %
8). In \S 12 we consider statistical mechanics. We thus arrive at a
systematic theory of evolving natural constants passing all the crucial
tests of Einstein's theory of general relativity and agreeing with the
empirical large number equalities (1.1), (1.2) or (1.4) when applied to
cosmology. We leave the exact numerical value $n$ introduced in the
generalized Dirac's large number hypothesis to be determined in the future
by fitting accurate relevant astrophysical or cosmological observations. For
the moment one may take $n=1$, the only integer in tne domain $0\,<n\,<2$,
if one likes, following Dirac's original large number hypothesis.

\section{Variational equations}

It is convenient (cf.(2.6) and (2.1)) to introduce the dimensionless new
variable
\begin{equation*}
\phi =G^{1/(2n)}\text{ \ , \ \ }\phi ^{2}=G^{1/n}\text{ \ \ (3.1)}
\end{equation*}%
\ so that (2.6) becomes
\begin{equation*}
\widetilde{I}_{g}=(16\pi )^{-1}\int (\phi ^{2}R_{\mu \nu }-w\phi _{;\mu
}\phi _{;\nu })g^{\mu \nu }\surd (-g)d^{4}x\text{ \ \ (3.2)}
\end{equation*}%
We obtain the variation as \
\begin{equation*}
\delta \widetilde{I}_{g}=(16\pi )^{-1}\int [-N^{\mu \nu }\delta g_{\mu \nu
}+2\Phi \delta \phi ]\surd (-g)d^{4}x\text{ \ \ \ \ \ (3.3)}
\end{equation*}%
with

\begin{eqnarray*}
\ N_{\alpha }^{\beta } &=&\phi ^{2}[R_{\alpha }^{\beta }-(1/2)R_{\sigma
}^{\sigma }\delta _{\alpha }^{\beta }]+(\phi ^{2})_{;\alpha }^{;\beta
}-(\phi ^{2})_{;\sigma }^{;\sigma }\delta _{\alpha }^{\beta } \\
&&-w[\phi _{;\alpha }\phi ^{; \beta }-(1/2)\phi _{;\sigma }\phi
^{;\sigma }\delta _{\alpha }^{\beta }]\ \ \ \ \ \ \ \ \
\text{(3.4)}\
\end{eqnarray*}%
(here we use the Palatini identity to express $\delta R_{\mu \nu }$ in terms
of the second covariant derivatives of $\delta g_{\alpha \beta }$ and
integrate by parts twice), and
\begin{equation*}
\Phi =R_{\sigma }^{\sigma }\phi +w\phi _{;\sigma }^{;\sigma }\text{ \ \ \ \
\ \ \ \ \ \ \ (3.5)}
\end{equation*}%
We note the identities (which can be shown from (3.3) by using an
infinitesmal transformation of coordinates or explicitly verified from (3.4)
and (3.5) by using the formula $(A^{\lambda })_{;\mu ;\nu }\ -(A^{\lambda
})_{;\nu ;\mu }=-$\ $A^{\sigma }R_{...\sigma \mu \nu }^{\lambda }$ with $%
\lambda =\nu $ summed), namely%
\begin{equation*}
N_{\alpha ;\beta }^{\beta }+\Phi \phi _{;\alpha }=0\text{ \ \ \ \ (3.6)}
\end{equation*}%
\ \

To calculate the variation of $\widetilde{I}_{m}$, (2.3a), we
follow closely Dirac's treatment for $I_{m}$,\ (E2.3a), as given
in his book \cite{Dirac2} on general theory of relativity. (Only
we use the letter $u$ instead of Dirac's $v$ to denote the
four-dimensional velocity) \
\begin{equation*}
\delta \widetilde{I}_{m}=-(1/2)\int \widetilde{T}^{\mu \nu }\delta g_{\mu
\nu }\surd (-g)d^{4}x-\int u_{\mu }\delta \widetilde{\text{$\wp $}}^{\mu
}d^{4}x\text{ \ \ \ (3.7)}
\end{equation*}%
with%
\begin{equation*}
\ \bigskip \widetilde{T}^{\mu \nu }=(g_{\alpha \beta }\widetilde{\text{$\wp $%
}}^{\alpha }\widetilde{\text{$\wp $}}^{\beta })^{-1/2}\widetilde{\text{$\wp $%
}}^{\mu }\widetilde{\text{$\wp $}}^{\nu }/\surd (-g)=\widetilde{\rho }%
_{m}u^{\mu }u^{\nu }\text{ \ \ (3.8)}
\end{equation*}%
by a change of notation that
\begin{eqnarray*}
u^{\mu } &=&(g_{\alpha \beta }\widetilde{\text{$\wp $}}^{\alpha }\widetilde{%
\text{$\wp $}}^{\beta })^{-1/2}\widetilde{\text{$\wp $}}^{\mu }\text{ \ \
(3.9)} \\
\text{\ }\widetilde{\rho }_{m}\surd (-g) &=&(g_{\alpha \beta }\widetilde{%
\text{$\wp $}}^{\alpha }\widetilde{\text{$\wp $}}^{\beta })^{1/2}\text{ \ \
\ \ \ \ (3.10)}
\end{eqnarray*}%
The second part of $\delta \widetilde{I}_{m}$, after substituting
\begin{equation*}
\delta \widetilde{\text{$\wp $}}^{\mu }=(\widetilde{\text{$\wp $}}^{\nu
}b^{\mu }-\widetilde{\text{$\wp $}}^{\mu }b^{\nu }),_{\nu }\text{ (3.11) \ \
so \ }\delta \widetilde{\text{$\wp $}}^{\mu },_{\mu }=0\text{ \ (3.12)}
\end{equation*}%
in compliance with the above constraint (2.3b), where $b^{\mu }$ is the
four-dimensional virtual displacement of the element of matter, becomes
after integrating by parts $\ $%
\begin{eqnarray*}
-\int u_{\mu }\delta \widetilde{\text{$\wp $}}^{\mu }d^{4}x &=&-\int
u_{\mu }(\widetilde{\text{$\wp $}}^{\nu }b^{\mu }-\widetilde{\text{$\wp $}}%
^{\mu }b^{\nu }),_{\nu }d^{4}x \\
&=&\int (u_{\mu },_{\nu }-u_{\nu },_{\mu })\widetilde{\text{$\wp $}}^{\nu
}b^{\mu }d^{4}x\ \ \ \ \ \ \ \ \text{(3.13)}\ \ \ \ \ \ \ \
\end{eqnarray*}%
The vanishing of this integral for an arbitrary $b^{\mu }$ gives the
geodesic equation of motion\ for the element of matter, namely
\begin{equation*}
\widetilde{\rho }_{m}u^{\nu }(u_{\mu },_{\nu }-u_{\nu },_{\mu })=\widetilde{%
\rho }_{m}u^{\nu }(u_{\mu ;\nu }-u_{\nu ;\mu })=\widetilde{\rho }_{m}u^{\mu
}u_{\nu ;\mu }=0\text{ \ (3.14)}
\end{equation*}%
the other part vanishing identically because we have from (3.9) the identity
$u^{\nu }u_{\nu }=1$ so $u^{\nu }u_{\nu ;\mu }=0$. Here we note the identies
similar to (3.6)%
\begin{equation*}
\widetilde{T}_{\nu ;\mu }^{\mu }=(u^{\mu }u_{\nu })_{;\mu }=(\widetilde{\rho
}_{m}u^{\mu })_{;\mu }u_{\nu }+\widetilde{\rho }_{m}u^{\mu }u_{\nu ;\mu }=0%
\text{ \ \ (3.15)}
\end{equation*}%
The identities (3.6) and (3.15) show that the system of variational
equations obtained from the comprehensive action principle $\delta
\widetilde{I}_{tot}=\delta \widetilde{I}_{g}+\delta \widetilde{I}_{m}=0$,
namely the gravitational field equations for $g_{\mu \nu }$ and $\phi $,%
\begin{eqnarray*}
N_{\alpha }^{\beta } &=&\phi ^{2}[R_{\alpha }^{\beta
}-(1/2)R_{\sigma }^{\sigma }\delta _{\alpha }^{\beta }]+(\phi
^{2})_{;\alpha }^{;\beta }-(\phi ^{2})_{;\sigma }^{;\sigma }\delta
_{\alpha }^{\beta }-w[\phi _{; \alpha }\phi ^{; \beta }-(1/2)\phi
_{;\sigma }\phi ^{;\sigma
}\delta _{\alpha }^{\beta }] \\
&=&-8\pi \widetilde{T}_{\alpha }^{\beta }=-8\pi \widetilde{\rho }%
_{m}u_{\alpha }u^{\beta }\text{ \ \ \ \ \ \ \ (3.16)}
\end{eqnarray*}%
\begin{equation*}
\text{ \ \ \ }\Phi =(w\phi _{;\sigma }^{;\sigma }+R_{\sigma }^{\sigma }\phi
)=0\text{ \ \ (3.17)}
\end{equation*}%
and the equations of motion (3.14) for the elements of matter are
compatiable but indeterminate. This is only natural for invariant action
integrals because the system of equations must allow $g_{\mu \nu }$ to
change with transformation of coordinates.

We note the following combination of (3.16) and (3.17) gives a particular
simple equation,
\begin{equation*}
N_{\sigma }^{\sigma }+\phi \Phi =(w/2-3)(\phi ^{2})_{;\sigma }^{;\sigma
}=-8\pi \widetilde{T}_{\sigma }^{\sigma }=-8\pi \widetilde{\rho }_{m}\text{
\ (3.18)}
\end{equation*}

It is interesting to write the field equations for $g_{\mu \nu }$ in a form
comparable with that in Einstein's theory involving a cosmological term.
Substituting (3.4) into (3.16) and dividing both sides by $\phi ^{2}$\ we
note that on the right-hand-side we have from (2.5) and (3.1)
\begin{equation*}
\widetilde{\rho }_{m}/\phi ^{2}=G\rho _{m}\text{ \ (3.19) or }\widetilde{T}%
_{\alpha }^{\beta }/\phi ^{2}=GT_{\alpha }^{\beta }\text{\ \ (3.20)}
\end{equation*}%
Hence we obtain
\begin{equation*}
R_{\alpha }^{\beta }-(1/2)R_{\sigma }^{\sigma }\delta _{\alpha }^{\beta
}+\Lambda _{\alpha }^{\beta }=-8\pi GT_{\alpha }^{\beta }\text{ \ \ (3.21)}
\end{equation*}%
where on the right $G$ is not a constant but varies as $t^{-n}$ while on the
left
\begin{eqnarray*}
\Lambda _{\alpha }^{\beta } &=&[(\phi ^{2})_{;\alpha }^{;\beta }-(\phi
^{2})_{;\sigma }^{;\sigma }\delta _{\alpha }^{\beta }]/\phi ^{2} \\
&&-w[\phi _{; \alpha }\phi ^{; \beta }-(1/2)\phi _{;\sigma
}\phi ^{;\sigma }\delta _{\alpha }^{\beta }]/\phi ^{2}\text{ \ (3.22)}
\end{eqnarray*}%
\ \ \ differs in its tensor character from Einstein's $\Lambda
\delta _{\alpha }^{\beta }=-\lambda \delta _{\alpha }^{\beta }$
and is originated from the variation of $G$\ , remembering (3.1).
In case $G=const$\ in a region, $\Lambda _{\alpha }^{\beta }$\
vanishes in the same region. (Here\ we use the capital Greek
$\Lambda _{\alpha }^{\beta }$ in (3.21) to compare with the
similar equation in Tolman's book \cite{Tolman} on relativity,
thermodynamics and cosmology which we write here with $G_{N}$
restored and
one index lowered%
\begin{equation*}
-8\pi G_{N}T_{\alpha }^{\beta }=R_{\alpha }^{\beta }-(1/2)R_{\sigma
}^{\sigma }\delta _{\alpha }^{\beta }+\Lambda \delta _{\alpha }^{\beta }
\end{equation*}%
The corresponding equation in Weinberg's book \cite{Weinberg} on
gravitation and cosmology is
\begin{equation*}
R_{\alpha }^{\beta }-(1/2)R_{\sigma }^{\sigma }\delta _{\alpha }^{\beta
}-\lambda \delta _{\alpha }^{\beta }=-8\pi G_{N}T_{\alpha }^{\beta }
\end{equation*}%
so $\lambda =-\Lambda $. To compare with the latter we should use $\lambda
_{\alpha }^{\beta }=-\Lambda _{\alpha }^{\beta }$) \

\section{Matter dominated cosmological model}

We can adopt as usual the Robertson-Walker metric because in
simplifying the metric to this form only considerations on
symmetry and freedom of coordinate transformation have been used,
but no use is made of the field equations. With $t$ denoting the
cosmic time and $r,\theta ,\varphi $\ the dimensionless co-moving
coordinates,
$ds^{2}=dt^{2}-R^{2}(t)\{dr^{2}/(1-kr^{2})+r^{2}d\theta
^{2}+r^{2}\sin ^{2}\theta d\varphi ^{2}\}$, $k=1,0,\textrm{or }-1$
all the equations of motions for the element of matter are
trivially satisfied by $u^{4}=u_{4}=1,u^{i}=u_{i}=0$ i.e. by being
at rest in the co-moving coordinates. From the modified constraint
(2.3b) we obtain, using dots for time derivatives,
\begin{equation*}
\{\widetilde{\rho }_{m}R^{3}[(1-kr^{2})^{-1/2}r^{2}\sin \theta ]\}^{\cdot }=0%
\text{ i.e. }\widetilde{\rho }_{m}=\widetilde{\rho }%
_{m}(t_{0})R^{3}(t_{0})/R^{3}\text{ \ \ (4.1)}
\end{equation*}%
where the constant $t_{0}$\ is really arbitrary but we often choose it to be
the time or the age of universe at present. Then (3.18) can be integrated to
give
\begin{equation*}
R^{3}(\phi ^{2})^{\cdot }=(3-w/2)^{-1}8\pi \widetilde{\rho }%
_{m}(t_{0})R^{3}(t_{0})t\text{ \ \ (4.2)}\ \ \ \ \
\end{equation*}%
Here the constant of integration additive to $t$\ is chosen to be zero so
that $t$ is counted since the big bang when $R=0.$ We know according to the
generalized large number hypothesis (2.1) that by (3.1) $\phi ^{2}\varpropto
t^{-1}$ in non-static homogeous cosmological model, hence (4.2) shows that $%
R\varpropto t$. Writing proportions as equalities
\begin{eqnarray*}
\phi ^{2} &=&t_{0}\phi ^{2}(t_{0})/t\text{ \ \ (4.3) } \\
\text{\ \ }R/R(t_{0}) &=&t/t_{0}\text{ (4.4) or }R=\beta _{ex}t\text{\ \
(4.5)}
\end{eqnarray*}%
$\beta _{ex}$ being the dimensionless proportional constant of the expanding
universe, we see that (4.2) is satisfied and gives
\begin{equation*}
8\pi \widetilde{\rho }_{m}(t_{0})/\phi ^{2}(t_{0})=8\pi G(t_{0})\rho
_{m}(t_{0})=(w/2-3)/t_{0}^{2}\text{ \ \ (4.6)}
\end{equation*}%
Substituting (4.3) and (4.4a,b) into (3.17), (3.16), we see all the field
equations are satisfied if we choose for our model with
\begin{equation*}
\text{\ \ \ }w=8(1+k/\beta _{ex}^{2})\text{ \ \ (4.7)}
\end{equation*}%
\ For example the field equation (3.17), namely \
\begin{eqnarray*}
R_{\mu }^{\mu }\phi +w\phi _{;\mu }^{;\mu } &=&6[\overset{\cdot \cdot }{R}%
/R+(\overset{\cdot }{R}^{2}+k)/R^{2}]\phi
\text{ }+w[\overset{\cdot \cdot }{\phi }+3(\overset{\cdot }{R}/R)\overset{%
\cdot }{\phi }] =0\ \ \ \ \ \ \ \ \ \ \ \ \ \text{(4.8)}
\end{eqnarray*}%
and the field equations $N_{1}^{1}=N_{2}^{2}=N_{3}^{3}=0$ which all coincide
into, cf.(3.16),
\begin{eqnarray*}
N_{1}^{1} &=&\phi ^{2}[-(\overset{\cdot }{R}^{2}+k)/R^{2}-2\overset{\cdot
\cdot }{R}/R]-(\phi ^{2})^{\cdot \cdot }-2(\phi ^{2})^{\cdot }\overset{\cdot
}{R}/R \\
+w\overset{\cdot }{\phi }^{2}/2 &=&0\text{ \ \ \ \ \ \ \ \ \ \ \ \ \ \ \ \
(4.9)\ }
\end{eqnarray*}%
both agree to determine $w$ as above$,$ while the equation%
\begin{eqnarray*}
N_{4}^{4} &=&\phi ^{2}[-3(\overset{\cdot }{R}^{2}+k)/R^{2}]-3(\phi
^{2})^{\cdot }\overset{\cdot }{R}/R \\
-w\overset{\cdot }{\phi }^{2}/2 &=&-8\pi \rho _{m}^{C}\text{ \ \ (4.10)}
\end{eqnarray*}%
gives a relation similar to (4.6) but with the factor $(w/2-3)$ replaced by $%
(w/8+3k/\beta _{ex}^{2})$ . These two factors, however, agree to be equal to
$(1+4k/\beta _{ex}^{2})$ by (4.7).\ \ Thus (4.3) and (4.5) are indeed the
solution of our non-static homogeneous cosmologic model, where (see
(4.6),(4.7))
\begin{equation*}
8\pi G(t_{0})\rho _{m}(t_{0})t_{0}^{2}=1+4k/\beta _{ex}^{2}=8\pi G\rho
_{m}t^{2},\text{ \ \ (4.11)}
\end{equation*}%
holds for arbitrary $t_{0}$ or $t$. Together with $4\pi \rho _{m}R^{3}/3=M$,
we obtain from (4.11)
\begin{equation*}
GM=(\beta _{ex}^{3}/6)(1+4k/\beta _{ex}^{2})c^{3}t\text{ \ \ (4.12)}
\end{equation*}%
in agreement with the large number relation (1.4) from where we started.
With (2.1) and (2.2), (4.12) becomes simply the conservation of $\widetilde{M%
}$. Also (4.11) shows by (2.1) that $\rho _{m}\varpropto t^{-(2-n)}$ which
suggests the domain for $n$ given in (2.1) above. We have from (3.22) and
(4.3)(4.4) , or cf. (4.9) and (4.10),
\begin{eqnarray*}
\Lambda _{1}^{1} &=&\Lambda _{2}^{2}=\Lambda _{3}^{3}=(1+k/\beta
_{ex}^{2})/t^{2} \\
\Lambda _{4}^{4} &=&(2-k/\beta _{ex}^{2})/t^{2}\text{ \ \ \ \ (4.13)}
\end{eqnarray*}%
We believe that the action integral $\widetilde{I}_{g}$ should be given once
for all, being independent of the problems as here treated. From (4.6) we
must choose $w>6$, and this excludes by (4.7) the case of $k=-1$ for $\beta
_{ex}^{2}<1$. For the same reason the case of $k=+1$ can occur only for $%
w\geq 16$. Our first preference is to choose\ $w=8$, thereby the case of
flat three dimensional space $k=0$ is determined. The second choice $w=16$,
which corresponds to the case $k=+1$ and $\beta _{ex}=1$, is far less
probable, we think. This question may be settled by noting the difference in
(4.13), i.e.
\begin{eqnarray*}
\Lambda _{4}^{4}/\Lambda _{1}^{1} &=&2\text{ for }w=8\text{ \ (4.13a)} \\
\Lambda _{4}^{4}/\Lambda _{1}^{1} &=&1/2\text{ for }w=16\text{ (4.13b)}
\end{eqnarray*}%
Of course a fresh analysis of the data of modern observation according to
the formulae of the present theory would be very important. Our cosmology
term varies inversely with the square of the cosmic time. In terms of the
Hubble function obtained from (4.4) (please note the different numerical
coefficient from Einstein's theory)%
\begin{equation*}
H(t)=\overset{\cdot }{R}(t)/R(t)=1/t\text{ \ \ (4.14)}
\end{equation*}%
both the age of universe $t_{0}$ and the values of our cosmological term at
present can be determined from the Hubble constant $H_{0}=H(t_{0})$ defined
by small cosmic redshifts as discussed in \S 9. This seems to give a natural
explaination for the order of magnitude of the cosmological constant at
present.

\section{Non-cosmological approximation}

\bigskip We are intrerested in examine whether our modified theory may pass
the crucial tests that support Einstein's theory of general relativity. Here
we shall consider in particular that refering to the advance of perihelion\
of Mercury. We need only the exterior solution for $g_{\mu \nu }$\ and $\phi
$\ good for, cosmologically speakng, a small region of space around the sun
and a short inteval $\Delta t$\ around $t_{0}$ say\ the present epoch. The
motion of the planet e.g. Mercury will then be obtained from the geodesic
equations of motion (3.4). We consider at first the cosmic background, i.e.
the cosmological model studied in \S 4. We write the dimensionless co-moving
coordinate in the Robertson-Walker metric in this section as $r_{R-W}$\ and
introduce a new variable $r$ of dimension length
\begin{equation*}
r=R(t_{0})r_{R-W}\text{ \ \ (5.1)}
\end{equation*}%
The metric becomes for example in the case of $k=0$
\begin{equation*}
\ ds^{2}=dt^{2}-[R^{2}(t)/R^{2}(t_{0})](dr^{2}+r^{2}d\theta ^{2}+r^{2}\sin
^{2}\theta d\varphi ^{2})\ \ \ \text{(5.2)}
\end{equation*}%
To the non-cosmological approximation we neglect $\Delta t$ against $t_{0}$
and so, by (4.4),
\begin{equation*}
R^{2}(t)/R^{2}(t_{0})=t^{2}/t_{0}^{2}=1+2\Delta t/t_{0}\doteq 1\text{ \ (5.3)%
}
\end{equation*}%
the space-time becomes flat, the background expansion being insignificant.
The non-cosmological approximation can simply be expressed as $%
t_{0}^{-1}\rightarrow 0,$ in the sense that terms involving $t_{0}$\ in the
denominator can be neglected. To this approximation $G$ (hence $\phi $)
assumes its quasistatic value, for example
\begin{equation*}
G(t)/G(t_{0})=(t/t_{0})^{-n}=1-n\Delta t/t_{0}\doteq 1\text{ \ \ (5.4)}
\end{equation*}%
This is consistent with $\Lambda _{\alpha }^{\beta
}=O(t_{0}^{-2})\rightarrow 0$ by (4.13) and $G\rho
_{m}=O(t_{0}^{-2})\rightarrow 0$ by (4.11), i.e. the background becomes the
vacuum and cosmological terms disappear. \ \

Now consider the sun as a local concentration of matter with density $\rho
(r)$ much higher than the cosmic background.. To the non-cosmological
approximation we can solve the system of gravitational field equations
(3.16) and (3.17) for $g_{\mu \nu }$\ and $\phi $\ outside the sun by the
quasistatic ($t/t_{0}=1+\Delta t/t_{0}=1$) exterior solution%
\begin{equation*}
\text{Exterior \ }\phi (t,r)=\phi (t_{0})=const\text{ \ \ (5.5) \ }R_{\mu
\nu }=0\text{ \ (5.6) }
\end{equation*}%
with the boundary condition at spatial infinity $g_{\mu \nu }$\ tending to,
according to (5.3), $\eta _{\mu \nu }$, that of flat space-time. Thus we
see, as far as exterior solution are concerned, our theory reduces to
Einstein's theory in the non-cosmological approximation. So our theory
passes the crutial test about the advance of perihelion.

\section{Electromagnetism for varying G}

In this section we shall go to find the modification for varying $G$ of the
action integral for the electromagnetic field and the additional action
integral for charged matter. We start from the large number equation (1.2).
According to our approach of replacing evolving law by cosmic combined
conservation law, for example see (2.4)(2.5) for $\widetilde{I}_{m}$ for
matter, which we consider it applicable also to the sun, the planet, even to
a mass point, so
\begin{equation*}
\widetilde{M}=G^{1+1/n}M,\text{ }\widetilde{m}_{p}=G^{1+1/n}m_{p},\text{ }%
\widetilde{m}_{e}=G^{1+1/n}m_{e}\text{ \ \ (6.1)}
\end{equation*}%
and the left hand side of (1.2) is a fixed constant. Assuming $c$ not
evolving, we conclude that $e^{2}/m_{e}\varpropto t\varpropto G^{-1/n}$, so
the cosmic combined charge\
\begin{equation*}
\text{ \ }\widetilde{e}=G^{1/2+1/n}e\text{ \ \ (6.2)}
\end{equation*}%
conserves in the long.

In Einstein's theory with constant $G,$ the action integral for the
electromagnetic field is (we follow again closely Dirac's book on general
theory of relativity)%
\begin{equation*}
I_{em}=-(16\pi )^{-1}\int F_{\mu \nu }F^{\mu \nu }\surd (-g)d^{4}x\text{ \
where }F_{\mu \nu }=\kappa _{\mu ;\nu }-\kappa _{\nu ;\mu }\text{ (E6.3a,b)}
\end{equation*}%
and the additional action integral\ for charged matter is (cf Dirac
(29.4)(29.1))%
\begin{equation*}
I_{q}=-\int \kappa _{\mu }\text{\^{J}}^{\mu }d^{4}x\text{ \ or }=-e\int
\kappa _{\mu }dx^{\mu }\text{ \ (E6.4a,b)}
\end{equation*}%
for continuous distribution of charges or for a point charge. Recalling the
rule of our modification mentioned before in connection with $I_{m}$\ and\ $%
\widetilde{I}_{m}$ , i.e. mutiplying $I$ with $G^{1+1/n}$\ and keep an eye
on the conservation law, we get naturally, starting from (6.2) and (E6.4b),
then (E6.4a) or (E6.3b), and finally (E6.3a), the following modified action
integrals%
\begin{equation*}
\widetilde{I}_{q}=-\widetilde{e}\int \widetilde{\kappa }_{\mu }dx^{\mu }%
\text{ with }\widetilde{\kappa }_{\mu }=G^{1/2}\kappa _{\mu }\text{ \
(6.4b,c)}
\end{equation*}%
\begin{equation*}
\widetilde{I}_{q}=-\int \widetilde{\kappa }_{\mu }\widetilde{\text{\^{\j}}}%
^{\mu }d^{4}x\text{ with }\widetilde{\text{\^{\j}}}^{\mu }=G^{1/2+1/n}\text{%
\^{\j}}^{\mu }\text{ (6.4a,d)}
\end{equation*}%
\begin{equation*}
\widetilde{F}_{\mu \nu }=\widetilde{\kappa }_{\mu {};\nu }-\widetilde{\kappa
}_{\nu ;\mu }\text{ }=\widetilde{\kappa }_{\mu },_{\upsilon }-\widetilde{%
\kappa }_{\nu },_{\mu }\text{\ \ (6.3b)}
\end{equation*}%
\begin{equation*}
\widetilde{I}_{em}=-(16\pi )^{-1}\int \phi ^{2}\widetilde{F}_{\mu \nu }%
\widetilde{F}^{\mu \nu }\surd (-g)d^{4}x\text{ \ (6.3a)}
\end{equation*}%
We give the variations of (6.3a)
\begin{eqnarray*}
\delta \widetilde{I}_{em} &=&\int [(-1/2)\widetilde{E}^{\mu \nu }\delta
g_{\mu \nu }+(4\pi )^{-1}(\phi ^{2}\widetilde{F}^{\mu \nu })_{;\nu }\delta
\kappa _{\mu } \\
&&+\Phi _{em}\delta \phi ]\surd (-g)d^{4}x\text{ \ (6.5a)}
\end{eqnarray*}%
with
\begin{equation*}
\widetilde{E}_{\mu }^{\nu }=\phi ^{2}[-(4\pi )^{-1}\widetilde{F}_{\mu \sigma
}\widetilde{F}^{\nu \sigma }+(16\pi )^{-1}\widetilde{F}_{\rho \sigma }%
\widetilde{F}^{\rho \sigma }\delta _{\mu }^{\nu }\text{ \ (6.5b)}
\end{equation*}%
$\ \ \ \ $%
\begin{equation*}
\ \Phi _{em}=-(8\pi )^{-1}\phi \widetilde{F}_{\mu \nu }\widetilde{F}^{\mu
\nu }\text{ \ (6.5c)}
\end{equation*}%
The variation of (6.4a) is%
\begin{equation*}
\delta \widetilde{I}_{q}=-\int [\widetilde{\text{\^{\j}}}^{\mu }\delta
\widetilde{\kappa }_{\mu }+\widetilde{\kappa }_{\mu }\delta \widetilde{\text{%
\^{\j}}}^{\mu }]d^{4}x\text{ \ \ (6.6)}
\end{equation*}%
In the comprehensive action principle $\delta \widetilde{I}_{tot}=0,$ $%
\widetilde{I}_{tot}=\widetilde{I}_{g}+\widetilde{I}_{m}+\widetilde{I}_{em}+%
\widetilde{I}_{q}$ we obtain from the coefficients of $\delta \widetilde{%
\kappa }_{\mu }$ the variational equation
\begin{equation*}
\widetilde{\text{\^{\j}}}^{\mu }=(4\pi )^{-1}(\phi ^{2}\widetilde{F}^{\mu
\nu })_{;\nu }\surd (-g)=(4\pi )^{-1}[\phi ^{2}\widetilde{F}^{\mu \nu }\surd
(-g)],_{\nu }\text{ \ (6.7)}
\end{equation*}%
Hence the identity
\begin{equation*}
\widetilde{\text{\^{\j}}}^{\mu },_{\mu }=0\text{, so }\int \widetilde{\text{%
\^{\j}}}^{4}dx^{1}dx^{2}dx^{3}\text{ conserves \ (6.8)}
\end{equation*}%
which reduces by (6.4d) to (6.2) for a point charge. In compliance with this
conservation law, we take in (6.6) as familiar in (3.11)%
\begin{equation*}
\delta \widetilde{\text{\^{\j}}}^{\mu }=(\widetilde{\text{\^{\j}}}^{\nu
}b^{\mu }-\widetilde{\text{\^{\j}}}^{\mu }b^{\nu }\text{),}_{\nu }\text{ \
(6.9)}
\end{equation*}%
After integrating by parts we obtain for th second part of (6.6)
\begin{equation*}
-\int \widetilde{\kappa }_{\mu }(\widetilde{\text{\^{\j}}}^{\nu }b^{\mu }-%
\widetilde{\text{\^{\j}}}^{\mu }b^{\nu }\text{),}_{\nu }d^{4}x=\int
\widetilde{F}_{\mu \nu }\widetilde{\text{\^{\j}}}^{\nu }b^{\mu }d^{4}x\text{
\ (6.10)}
\end{equation*}%
For the elements of charged matter the equations of motion obtained from the
coefficients of $b^{\mu }$ , by a change of notation similar to (3.9)(3.10)
\begin{equation*}
\widetilde{\text{\^{\j}}}^{\nu }=\widetilde{\rho }_{e}u^{\nu }\surd (-g)%
\text{ , \ }\widetilde{\rho }_{e}=G^{1/2+1/n}\rho _{e}\text{ \ (6.11a,b)}
\end{equation*}%
contains a Lorentz force term%
\begin{equation*}
\widetilde{\rho }_{m}u^{\mu }u_{\nu ;\mu }+\widetilde{F}_{\mu \nu }%
\widetilde{\rho }_{e}u^{\nu }=0\text{ \ (6.12)}
\end{equation*}%
Dirac, in his book, uses $\sigma $\ instead of our $\rho _{e}$\ for the
charge density. The variational equations for $g_{\mu \nu }$\ and $\phi $\
are
\begin{equation*}
N_{\mu }^{\nu }=-8\pi \widetilde{T}_{\mu }^{\nu }-8\pi \widetilde{E}_{\mu
}^{\nu }\text{, }\Phi =-8\pi \Phi _{em}\text{ (6.13a,b)}
\end{equation*}%
The effect of the factor$\ \phi ^{2}$ in (6.3a) and (6.7) is familiar from
Maxwell's theory; as seen from (6.7) in the case of Galelian metric $\phi
^{2}$ plays the role of relative dielectric constant $\varepsilon _{\phi }$
and $(\phi ^{2})^{-1}$\ plays the role of relative magnetic susceptibility $%
\mu _{\phi }$.\ Hence $\varepsilon _{\phi }\mu _{\phi }=1$ which is
consistent with our assumption that $c$\ does not evolve.

\section{Geometric optics}

We shall study in this section the propagation of light or electromagnetic
wave in space time where the metric and $\phi $ are given. As is well known
in Einstein's theory of general relativity that light moves along a null
geodesic, and this can be regarded as a good approximation of geometric
optics to wave optics for the electromagnetic equations in the Riemanian
space-time. In this section we shall show that this is also true\ for the
theory with varying $G$, following closely the smilar treatment$^{\text{5}}$
of the present author for constant $G$.

From (6.3b) where the covariant derivative can be replaced by ordinary
partial derivatives, we can eliminate the cosmic combined potentials and
obtain
\begin{equation*}
\widetilde{F}_{\mu \nu },_{\lambda }+\widetilde{F}_{\nu \lambda },_{\mu }+%
\widetilde{F}_{\lambda \mu }{}_{\nu }=0\text{ \ (7.1)}
\end{equation*}%
From (6.7) where charge-current is absent we have
\begin{equation*}
\lbrack \phi ^{2}\widetilde{F}{}^{\mu \nu }\surd (-g)],_{\nu }=[\phi
^{2}g^{\mu \alpha }g^{\nu \beta }\widetilde{F}_{\alpha \beta }\surd
(-g)],_{\nu }=0\text{ \ \ (7.2)}
\end{equation*}%
Among the four equations in the set (7.2) there is one differential identity
namely the ordinary divergence of its left hand side vanishes identically.
The same is true for the four equations in the set (7.1) which can be
written in a form similar to (7.2) by introducing the dual of $\widetilde{F}%
_{\mu \nu }$ defined by
\begin{equation*}
\widehat{\widetilde{F}}^{41}=\widetilde{F}_{23}\text{ ; }\widehat{\widetilde{%
F}}^{23}=\widetilde{F}_{41}\text{ ; }1,2,3\text{ cyclic \ (7.3)}
\end{equation*}%
Hence the set (7.1), (7.2) contains six independent equations linear in the
six dependent variables $\widetilde{F}_{\mu \nu }$. For light or
electromagnetic waves, we need to consider solutions of the form of waves
\begin{equation*}
\widetilde{F}_{\mu \nu }=\widetilde{f}_{\mu \nu }\sin S\text{ \ \ (7.4)}
\end{equation*}%
where the phase $S$\ varies quickly in space or time at the scale of a
wavelength or a period of light, but the amplitudes $\widetilde{f}_{\mu \nu
} $ vary little at such scale, because the multipliers in (7.2) only vary at
a scale much much larger. In contrast to the fast variable $S$\ we call $%
\widetilde{f}_{\mu \nu }$, $\phi ,g_{\rho \sigma }$\ all\ slow variables.
Neglecting the derivatives of the slow variables in comparison with those of
the fast variable, we obtain from (7.1) and (7.2) after removing the common
factor $\cos S$\ and writing $S,_{\lambda }=s_{\lambda }$ the equations\
\begin{equation*}
\widetilde{f}_{\mu \nu }s_{\lambda }+\widetilde{f}_{\nu \lambda }s_{\mu }+%
\widetilde{f}_{\lambda \mu }s_{\nu }=0\text{ \ \ (7.5)}
\end{equation*}%
\begin{equation*}
\lbrack \phi ^{2}g^{\mu \alpha }g^{\nu \beta }\widetilde{f}_{\alpha \beta
}\surd (-g)]s_{\nu }=0\text{\ i.e. }s^{\beta }\widetilde{f}_{\lambda \beta
}=0\text{\ \ \ (7.6)}
\end{equation*}%
where $s^{\beta }=g^{\nu \beta }s_{\nu }$.\ In (7.6) the second equation
follows from the first by contracting the latter with $g_{\lambda \mu }$ and
removing the common factors $\phi ^{2}$\ and $\surd (-g).$\ These algebraic
equations being homogeneous and linear in the amplitudes $\widetilde{f}_{\mu
\nu }$, the condition for non-trivial solution is the vanishing of the
determinant $s_{\lambda }s^{\lambda }=0$. More simply\ this can be obtained
by contracting (7.5) with $s^{\lambda }$\ and using (7.6) to obtain $%
\widetilde{f}_{\mu \nu }s_{\lambda }s^{\lambda }=0.$ So at any point where
there is light we must have%
\begin{equation*}
s_{\lambda }s^{\lambda }=g^{\lambda \mu }s_{\lambda }s_{\mu }=0\ \ \text{%
(7.7)}
\end{equation*}%
Let a real displacement along the light path be denoted by $dx^{\mu }$,
alomg the light path (7.7) always hold, so \
\begin{equation*}
d(g^{\lambda \mu }s_{\lambda }s_{\mu })=s_{\lambda }s_{\mu }g^{\lambda \mu
},_{\nu }dx^{\nu }+2s^{\nu }ds_{\nu }=0\text{ \ \ (7.8)}
\end{equation*}%
Let a virtual displacement on the surface of constant phase i.e. $S=const$.
be denoted by $\delta x^{\nu }$. So we have%
\begin{equation*}
0=\delta S=S,_{\nu }\delta x^{\nu }=s_{\nu }\delta x^{\nu }\text{ \ (7.9) \ }
\end{equation*}%
Since these two displacement vectors are orthogonal to each other, we have
also%
\begin{equation*}
g_{\mu \nu }dx^{\mu }\delta x^{\nu }=0\text{ \ \ (7.10)}
\end{equation*}%
On comparing the last two equations we see that the coefficients of $\delta
x^{\nu }$\ there must be proportional. As is well known, we can choose the
parameter $p$ for the path of light $x^{\mu }=x^{\mu }(p)$ such that (7.10)
agrees with (7.9) and (7.8) is satisfied\ by the following two set of first
order differential equations for the light paths or rays.
\begin{equation*}
g_{\mu \nu }(dx^{\mu }/dp)=s_{\nu },\text{ }ds_{\nu }/dp=-(1/2)g^{\lambda
\mu },_{\nu }s_{\lambda }s_{\mu }\text{ \ (7.11a,b)}
\end{equation*}%
In Riemanian geometry, (7.11a,b) is the standard form for the differential
equations of null geodesics, the two sets of fiest order equations being
completely equivalent to the following one set of second order equations by
eliminating $s_{\nu }$ with the help of (7.11a),
\begin{equation*}
d^{2}x^{\mu }/dp^{2}+\Gamma _{\rho \sigma }^{\mu }(dx^{\rho }/dp)(dx^{\sigma
}/dp)=0\text{ \ (7.12)}
\end{equation*}%
and the equation (7.7) being equivalent to the null condition
\begin{equation*}
0=g^{\mu \nu }s_{\mu }s_{\nu }=g^{\mu \nu }g_{\mu \rho }g_{\nu \sigma
}(dx^{\rho }/dp)(dx^{\sigma }/dp)=ds^{2}/dp^{2}\text{ \ (7.13)}
\end{equation*}

As we have shown that light path in our theory as in Einstein's theory is
the null geodesic and the metric outside the sun to the non-cosmological
approximation is the same as that in Einstein's theory, so the crucial test
about the deflection of light by the sun is also passed by our theory as
good as by Einstein's theory.

\section{Quantum mechanics}

The fundamental natural constant in quantum mechanics is the Planck constant
$\hslash $. In order to decide how it evolves, we need some action integral
that contains it. It seems most reliable to consider the action integral for
Dirac's relativistic theory of electron, because the generalization of
Dirac's equation to curved space-time i.e. to general relativity has been
considered by many early authors. Here we follow Schrodinger's treatment$\ $%
as summarized in my earlier paper \cite{Peng} with minor changes
in the notation (the electromagnetic potentials $A_{\mu }$\ there
is here written
as $\kappa _{\mu }$ and the companion $\phi $\ there of the wave function $%
\psi $\ is here written as $\chi $ ). We consider hydrogen like atoms for
simplicity. With constant $G$ we have%
\begin{eqnarray*}
I_{D} &=&\int \{(1/2)\chi G^{\mu }[(i\hslash \partial /\partial x^{\mu
}+e\kappa _{\mu })\psi ] \\
&&+(1/2)[(-i\hslash \partial /\partial x^{\mu }+e\kappa _{\mu })\chi ]G^{\mu
}\psi -\chi m\psi \}\surd (-g)d^{4}x\text{ \ (8.1)}
\end{eqnarray*}%
where the four general gamma $4\times 4$ matrices $G^{\mu }$\ depend only on
$g_{\mu \nu }$\ satisfying
\begin{equation*}
G^{\mu }G^{\nu }+G^{\nu }G^{\mu }=2g^{\mu \nu }\text{ \ (8.2)}
\end{equation*}%
and $\chi $\ and $\psi $\ are respectively $1\times 4$\ and $4\times 1$\
matrices so that the chain product $\chi G^{\mu }\psi $ is a number. Here $%
I_{D}$\ for the hydrogen atom includes the part previously denoted by $I_{q}$
. We have, cf (6.4a), for Dirac's electron
\begin{equation*}
\text{\^{\j}}^{\mu }=-e\chi G^{\mu }\psi \surd (-g)\text{ \ \ (8.3)}
\end{equation*}%
\ the ordinary divergence of which vanishes as a consequence of the
variational equations with respect to $\chi $\ and $\psi $, i.e. the Dirac
equation $\delta I_{D}/\delta \chi =0$ and its companion $\delta
I_{D}/\delta \psi =0$\ in\ curved space-time.
\begin{equation*}
0=\chi \frac{\delta I_{D}}{\delta \chi }-\frac{\delta I_{D}}{\delta \psi }%
\psi =[\chi G^{\mu }\psi \surd (-g)],_{\mu }\text{ \ (8.4)}
\end{equation*}%
The variation of $I_{D}$\ with respect to $g_{\alpha \beta }$\ can be
obtained from the dependence of \ $G^{\mu }$ on $g_{\alpha \beta }$\ by
using the formula derived from (8.2)
\begin{equation*}
\partial G^{\mu }/\partial g_{\alpha \beta }=-(1/4)(G^{\alpha }g^{\beta \mu
}+G^{\beta }g^{\alpha \mu })\text{ \ (8.5)}
\end{equation*}

According to our rule the modified action integral for varying $G$\ is
\begin{eqnarray*}
\widetilde{I}_{D} &=&\int \{(1/2)\chi G^{\mu }[(i\widetilde{\hslash }%
\partial /\partial x^{\mu }+\widetilde{e}\widetilde{\kappa }_{\mu })\psi ] \\
&&+(1/2)[(-i\widetilde{\hslash }\partial /\partial x^{\mu }+\widetilde{e}%
\widetilde{\kappa }_{\mu })\chi ]G^{\mu }\psi -\chi \widetilde{m}_{e}\psi
\}\surd (-g)d^{4}x\text{ \ (8.6)}
\end{eqnarray*}%
This confirms our previous relations (6.1)\ for $\widetilde{m}_{e}$ and
(6.2) (6.4c) for $\widetilde{e}$ $\widetilde{\kappa }_{\mu }$. It shows
further that the cosmic combined Planck constant
\begin{equation*}
\widetilde{\hslash }=G^{1+1/n}\hslash \text{ \ \ (8.7)}
\end{equation*}%
conserves in the long.

Meanwhile we call attention to the expression of cosmic combined Coulomb
potential $\widetilde{\kappa }_{4}$\ due to a point charge, say $Z\widetilde{%
e}$\ which, in the case of Galelian metric, is not $Z\widetilde{e}/r$ but is
$\widetilde{\kappa }_{4}=(\phi ^{2})^{-1}(Z\widetilde{e}/r)=Z\widetilde{e}%
/(\varepsilon _{\phi }r)$\ by solving (6.7) with $\mu =4$ . Using (6.2) and
(3.1), we verify that the Coulomb potential, only correctly written this
way, evolves like (6.4c)
\begin{eqnarray*}
\widetilde{\kappa }_{4} &=&G^{1/2}\kappa _{4}=G^{1/2}Ze/r\text{ } \\
\text{so \ }\widetilde{e}\widetilde{\kappa }_{4} &=&G^{1+1/n}e\kappa
_{4}=G^{1+1/n}[Ze^{2}/r]\text{ \ \ (8.8)}
\end{eqnarray*}%
Hence the fine structure constant defined by (here $c$ being restored)
\begin{equation*}
\widetilde{\alpha }=\widetilde{e}^{2}/(\varepsilon _{\phi }\widetilde{%
\hslash }c)=e^{2}/(\hslash c)=\alpha \text{ \ \ (8.9)}
\end{equation*}%
remains unchanged like $c$ during the evolution of the universe.

We are interested in the atomic spectra (of hydrogen say) for
light coming from extra-galactic nebulae or from the sun. To
obtain the energy levels by quantum mechanics\ we shall follow
closely the treatment in my earlier paper \cite{Peng} in
Einstein's theory. We use the local Galelian metric obtained by a
coordinate transformation, which can always be done if we treat
the metric $g_{\mu \nu }$ as constants, assuming their value at
the position of the atom at the time of emission or absorption of
the light quantum. For simplicity consider the diagonal case
\begin{equation*}
g_{ii}=-(\lambda _{i})^{2},i=1,2,3,\text{ }g_{44}=+(\lambda _{4})^{2},\text{
}g_{\mu \nu }=0\text{ }(\mu \neq \nu )\text{ \ (8.10)}
\end{equation*}%
Then we have%
\begin{equation*}
G^{i}=(\lambda _{i})^{-1}\gamma ^{i},i=1,2,3,\text{ }G^{4}=(\lambda
_{4})^{-1}\gamma ^{4}\text{ \ (8.11)}
\end{equation*}%
where $\gamma ^{\mu }$ are the $4\times 4$ gamma matrices familiar for
special relativity, defined by
\begin{equation*}
\gamma ^{\mu }\gamma ^{\nu }+\gamma ^{\nu }\gamma ^{\mu }=2\eta ^{\mu \nu }%
\text{ \ \ (8.12)}
\end{equation*}%
Then Dirac's equation as obtained from (8.6) by varying $\chi $ becomes
\begin{equation*}
\{\gamma ^{4}(\lambda _{4})^{-1}[i\widetilde{\hslash }(\partial /\partial t)+%
\widetilde{e}\widetilde{\kappa }_{4}]+\gamma ^{j}(\lambda _{j})^{-1}i%
\widetilde{\hslash }(\partial /\partial x^{j})-\widetilde{m}_{e}\}\psi =0%
\text{ \ (8.13)}
\end{equation*}%
where the vector potential vanishes and the scalar potential $\widetilde{%
\kappa }_{4}$\ due to the nucleus at rest at the origin is to be obtained by
solving (6.7) with $\mu =4,$ namely
\begin{equation*}
\lbrack \phi ^{2}\widetilde{F}^{4\nu }\surd (-g)],_{\nu }=4\pi (Z\widetilde{e%
})\delta (x^{1})\delta (x^{2})\delta (x^{3})\text{ \ (8.14)}
\end{equation*}%
In the region of atomic scale $\phi $ may also be treated as constant, so
the left hand side of this equation becomes simply $\phi
^{2}g^{44}g^{jj}\lambda _{1}\lambda _{2}\lambda _{3}\lambda _{4}(\widetilde{%
\kappa }_{4}),_{jj}$ summed over $j$. It is convenient to introduce the
scaled variables separately
\begin{equation*}
y^{j}=\lambda _{j}x^{j}\text{ so }\delta (y^{j})=\delta (x^{j})/\lambda _{j}%
\text{ }j=1,2,3\text{ not summed \ (8.15)}
\end{equation*}%
Then (8.14) becomes in the scaled variables
\begin{equation*}
\nabla _{y}^{2}(\widetilde{\kappa }_{4})=-\lambda _{4}(4\pi Z\widetilde{e}%
/\phi ^{2})\delta (y^{1})\delta (y^{2})\delta (y^{3})\text{ \ \ (8.16)}
\end{equation*}%
the solution being proportional to the inverse radial distance $r_{y}$\ of
the scaled $y$'s
\begin{equation*}
\widetilde{\kappa }_{4}=\lambda _{4}(Z\widetilde{e}/\phi ^{2})r_{y}^{-1}%
\text{ \ (8.17)}
\end{equation*}%
With the help of (8.17), the equation (8.13) can be written in the scaled
variables
\begin{equation*}
\{\gamma ^{4}[(\lambda _{4})^{-1}i\widetilde{\hslash }(\partial /\partial
t)+Z\widetilde{e}^{2}\phi ^{-2}r_{y}^{-1}]+\gamma ^{j}[i\widetilde{\hslash }%
(\partial /\partial y^{j})]-\widetilde{m}\}\psi =0\text{ (8.18)}
\end{equation*}%
or after removing the common factor $G^{1+1/n}$ (cf.(8.3) (6.1) (6.2) (3.1))
\begin{equation*}
\{\gamma ^{4}[(\lambda _{4})^{-1}i\hslash (\partial /\partial
t)+(Ze^{2})r_{y}^{-1}+\gamma ^{j}[i\hslash (\partial /\partial
y^{j})]-m\}\psi =0\text{ (8.19)}
\end{equation*}%
The energy value including the rest energy defined by
\begin{equation*}
i\widetilde{\hslash }(\partial /\partial t)\psi =\widetilde{E}\psi \text{ \
(8.20) \ \ }i\hslash (\partial /\partial t)\psi =E\psi \text{ \ (8.21)}
\end{equation*}%
is easily obtained by comparing with that of Dirac's equation in his book on
quantum mechanics.. Using the quantum numbers $n_{r}$ and $j$ we obtain
(with $c$\ restored and using (8.5) for $\lambda _{4}$)
\begin{equation*}
\widetilde{E}_{n_{r},,j}/\widetilde{m}%
c^{2}=E_{n_{r},j}/mc^{2}=(g_{44})^{1/2}\{1+\frac{Z^{2}\alpha ^{2}}{(n_{r}+%
\sqrt{j^{2}-Z^{2}\alpha ^{2}})^{2}}\}^{-1/2}\text{ (8.22)}
\end{equation*}%
For a transition between two quantum states $A$ and $B$ with quantum numbers
$n_{r}(A),j(A)$ and $n_{r}(B),j(B)$ the frequency $\nu _{A\rightarrow B}$ of
the atomic spectral line is
\begin{equation*}
\nu _{A\rightarrow B}=(\widetilde{E}_{A}-\widetilde{E}_{B})/\widetilde{h}%
=(E_{A}-E_{B})/h=(g_{44})^{1/2}(\nu _{A\rightarrow B})_{QM}\text{\ (8.23)}
\end{equation*}%
where the ordinary quantum mechanical value of the frequency is given by
\begin{equation*}
(\nu _{A\rightarrow B})_{QM}=\frac{c}{2\pi \lambdabar _{c}}\{1+\frac{%
Z^{2}\alpha ^{2}}{(n_{r}+\sqrt{j^{2}-Z^{2}\alpha ^{2}})^{2}}\}^{-1/2}\
|_{B}^{A}\text{ (8.24)}
\end{equation*}%
Since both the Compton wave length $\lambdabar_{c}=\hslash /mc=\widetilde{%
\hslash }/\widetilde{m}c=\widetilde{\lambdabar}_{c}$ and the fine structure
constant $\alpha =\widetilde{\alpha }$ remain unchanged, we see from (8.24)
that the frequency of the atomic spectral line does not change during the
evolution. Also the Rydberg constant $R_{\infty }=mc^{2}\alpha ^{2}/(4\pi
\hslash )=$ $\widetilde{R}_{\infty }$ remains unchanged. So does the mass
ratio e.g. $m_{p}/m_{e}=\widetilde{m}_{p}/\widetilde{m}_{e}$. The relation
(8.23), it seems, can be generalized to all energy levels of complex atoms
and molecules.

We have shown in (8.22) that the gravitational effect on the frequency of
the spectral line during emission or absorption in our theory is given by
the factor $(g_{44})^{1/2}$ , which is the same in our theory as in
Einstein's theory to the non-cosmological approximation. So the crutial test
about the gravitational red shift of the sun is also passed by our theory as
by Einstein's theory.

We note in the Robertson-Walker metric using cosmic time with $g_{44}=1$ the
frequency of the spectral lines emitted anywhere at any time is by (8.23)
always the same.

\section{Hubble relation}

In this section we shall investigate the relation between the
cosmic red shift of the spectral line received from a cosmic
distant source and the distance of that source from an observer
here at present. We use the Robertson-Walker metric (5.2), so the
frequency of the spectral lines for the\ distant and nearby object
are the same by (8.23), namely $\nu _{QM}$ just when they are
emitted. But there is a change of the frequency of the spectral
line of the distant source during its propagating towards the
receiver according to (7.11b).

We denote quantities referring to the distant source by the suffix $s$ and
those referring to the receiver or observer here at present by the suffix $0$%
. For simplicity we use geometric optics and radial rays, putting
$d\theta =d\varphi =s_{\theta }=s_{\varphi }=0$, the phase of the
spherical waves
being, as usual, $S=2\pi (\nu t-r/\lambda )=2\pi \nu (t-r/\beta )$, with $%
s_{4}=2\pi \nu ,s_{r}=-2\pi /\lambda $. The null condition gives the
velocity of light $\beta $ for the metric (5.2)
\begin{equation*}
\lambda \nu =\beta =R(t_{0})/R(t)\text{ \ (9.1)}
\end{equation*}%
So at the time $t_{s}$ the wavelength of the spectral line emitted by the
source is
\begin{equation*}
\lambda _{s}\nu _{QM}=R(t_{0})/R(t_{s})\text{ \ }(t_{s}<t_{0})\text{\ \ (9.2)%
}
\end{equation*}%
For the metric (5.2) , being homogeneous, the wave length of this spectra
line keeps constant by (7.11b) during the propagation, till the line being
recieved and compared at time $t_{0}$ with the same line emitted near by
with
\begin{equation*}
\lambda _{0}\nu _{QM}=1\text{ \ (9.3)}
\end{equation*}%
The cosmic redshift $z$ is defined by $z=(\lambda _{s}-\lambda _{0})/\lambda
_{0}$, so dividing (9.2) by (9.3) we have
\begin{equation*}
1+z=\lambda _{s}/\lambda _{0}=R(t_{0})/R(t_{s})\text{ \ \ (9.4)}
\end{equation*}%
The optical distance $\Delta $\ the light propagated is obtained by
integrating $dr$ from $t_{s}$ to $t_{0}$%
\begin{equation*}
\Delta =\int_{s}^{0}dr=\int_{t_{s}}^{t_{0}}\frac{R(t_{0})}{R(t)}dt\text{ \
(9.5)}
\end{equation*}%
If this is much larger compared to the linear dimension of the source the
inverse square law holds for the intensity and can be used to measure the
distance.

For matter dominated cosmological model we have $R(t_{0})/R(t)=t_{0}/t$, so
from (9.5) and (9.4) we obtain
\begin{equation*}
\Delta =t_{0}\log (t_{0}/t_{s})=t_{0}\log (1+z)\text{ \ \ (9.6)}
\end{equation*}%
This becomes the Hubble relation%
\begin{equation*}
z=H_{0}\Delta \text{ with }H_{0}=(t_{0})^{-1}\text{ for }z<<1\text{ \ \ (9.7)%
}
\end{equation*}%
but for larger $z$ we have from (9.6)%
\begin{equation*}
z=\exp (H_{0}\Delta )-1\text{ \ \ (9.8)}
\end{equation*}

\section{Electromagnetic radiation}

In the next section we shall supplement the matter dominated universe with
electromagnetic radiation. Owing to the appearance of the factor $\phi ^{2}$
in the action (6.3a) we need study the electromagnetic waves a bit more in
order to obtain the expressions corresponding to (6.5b) and (6.5c)\ for
electromagnetic radiation.

We write at first the electromagnetic wave equations in a form familiar from
\ Maxwell's theory. Free from source (6.7) splits into
\begin{eqnarray*}
div\overrightarrow{D} &=&0,\text{ \ (10.1) }\overrightarrow{D}=[\phi
^{2}\surd (-g)](\widetilde{F}^{41},\widetilde{F}^{42},\widetilde{F}^{43})%
\text{ \ (10.1a) \ } \\
curl\overrightarrow{H}-\partial D/\partial t &=&0\text{ \ \ (10.2) }%
\overrightarrow{H}=[\phi ^{2}\surd (-g)](\widetilde{F}^{23},\widetilde{F}%
^{31},\widetilde{F}^{12})\text{ \ (10.2a)}
\end{eqnarray*}%
The other pair of Maxwell's equations come from (6.3b), from which we obtain
$\widetilde{F}_{\mu \nu },_{\lambda }+\widetilde{F}_{\lambda \mu },_{\nu }+%
\widetilde{F}_{\nu \lambda },_{\mu }=0$ which splits into
\begin{eqnarray*}
div\overrightarrow{B} &=&0\text{ \ (10.3) }\overrightarrow{B}=(\widetilde{F}%
_{23},\widetilde{F}_{31},\widetilde{F}_{12})\text{ \ (10.3a)} \\
curl\overrightarrow{E}+\partial \overrightarrow{B}/\partial t &=&0\text{ \
(10.4) }\overrightarrow{E}=(\widetilde{F}_{14},\widetilde{F}_{24},\widetilde{%
F}_{34})\text{ \ (10.4a)}
\end{eqnarray*}%
Comparing (6.5c) with (6.3a), the integrand of the latter being equal to two
times $(\overrightarrow{B}\cdot \overrightarrow{H}-\overrightarrow{E}\cdot
\overrightarrow{D})$, we obtain
\begin{equation*}
\Phi _{em}=-(\overrightarrow{B}\cdot \overrightarrow{H}-\overrightarrow{E}%
\cdot \overrightarrow{D})/[4\pi \phi \surd (-g)]\text{ \ (10.5)}
\end{equation*}%
For the Robertson-Walker metric with $k=0,$ it is convenient to use
Cartesian coordinates to write
\begin{equation*}
ds^{2}=dt^{2}-f^{2}(dx^{2}+dy^{2}+dz^{2})\text{ with }f=R(t)/R(t_{0})\text{
\ (10.6)}
\end{equation*}%
Then we have from (10.1a) to (10.4a) the slowly varying permittences $%
\varepsilon $ and $\mu $
\begin{eqnarray*}
\overrightarrow{D} &=&\varepsilon \overrightarrow{E},\text{ \ \ (}%
\varepsilon =[\phi ^{2}f^{3}]f^{-2}\text{) \ \ \ (10.7a)} \\
\text{ }\overrightarrow{H} &=&\mu ^{-1}\overrightarrow{B},\text{ (}\mu
^{-1}=[\phi ^{2}f^{3}]f^{-4}\text{) \ (10.7b)}
\end{eqnarray*}

We treat the electromagnetic radiation as superposition of plane
electromagnetic waves. For a plane electromagnetic wave propagating along
the direction denoted by the unit vector $\overrightarrow{s}$, the only fast
variable is the phase $S=\overrightarrow{r}\cdot \overrightarrow{s}-\beta t$
where\ $\beta =1/\sqrt{\varepsilon \mu }$, like $\varepsilon $\ and $\mu $,
vary slowly in time and will be treated as constants. Then (we follow here
Born and Wolf's treatment$^{\text{6}}$) (10.2) and (10.4) becomes
\begin{equation*}
\overrightarrow{E}=-\sqrt{\mu /\varepsilon }(\overrightarrow{s}\times
\overrightarrow{H}),\overrightarrow{H}=\sqrt{\varepsilon /\mu }(%
\overrightarrow{s}\times \overrightarrow{E})\text{ (10.8)}
\end{equation*}%
Hence $\overrightarrow{E},\overrightarrow{H},$and $\overrightarrow{s}$ form
a right-handed orthogonal triad, and we have
\begin{equation*}
\varepsilon E^{2}=\mu H^{2}\text{ or }\overrightarrow{D}\cdot
\overrightarrow{E}-\overrightarrow{H}\cdot \overrightarrow{B}=0\text{ \
(10.9)}
\end{equation*}%
So by (10.5) we have $\Phi _{em}=0$\ for a plane electromagnetic
wave, and by superposition of plane waves we obtain for
electromagnetic radiation
\begin{equation*}
\Phi _{em}=0\text{ (electromagnetic radiation) \ (10.10) \ }
\end{equation*}

For electromagnetic field where charge-current density vanishes we
have the differential identity,
\begin{equation*}
\widetilde{E}_{\mu ;\nu }^{\nu }+\Phi _{em}\phi _{;\mu }=0\text{ \ \ (10.11)}
\end{equation*}%
which can easily be verified from (6.5b) and (6.5c). Where charge-current
density does not vanish, this identity should be replaced by
\begin{equation*}
\widetilde{T}_{\mu ;\nu }^{\nu }+\widetilde{E}_{\mu ;\nu }^{\nu }+\Phi
_{em}\phi _{;\mu }=0\text{ \ \ (10.12)}
\end{equation*}%
so that the terms involving Lorentz's force from $\widetilde{T}_{\mu ;\nu
}^{\nu }$ and $\widetilde{E}_{\mu ;\nu }^{\nu }$ cancel. For electromagnetic
radiation (10,11) simplifies by (10.10) to
\begin{equation*}
\widetilde{E}_{\mu ;\nu }^{\nu }=0\text{ (electromagnetic radiation) \ (10.13)%
}
\end{equation*}%
For homogeneous and isotropic electromagnetic radiation the superposition of
plane waves propagating in all directions leaves only the diagonal elements
of $\widetilde{E}_{\mu }^{\nu }$\ non-vanishing Since from (6.5b) we have
zero trace before the superposition, we conclude that
\begin{equation*}
\widetilde{E}_{1}^{1}=\widetilde{E}_{2}^{2}=\widetilde{E}_{3}^{3}=-%
\widetilde{p}_{r}=-\widetilde{\rho }_{r}/3,\text{ }\widetilde{E}_{4}^{4}=%
\widetilde{\rho }_{r}\text{ (10.14)}
\end{equation*}%
It is well known that with (10.14) and the metric (10.6) that (10.13) can be
integrated to obtain
\begin{equation*}
\widetilde{\rho }_{r}=\widetilde{\rho }_{r}(t)=\widetilde{\rho }%
_{r}(t_{0})R^{4}(t_{0})/R^{4}\text{ \ (10.15)}
\end{equation*}

\section{Matter plus electromagnetic radiation}

We now consider homogeneous cosmology including matter and electromagnetic
radiation, $\widetilde{I}_{tot}=\widetilde{I}_{g}+\widetilde{I}_{m}+%
\widetilde{I}_{em}$ and write according to our theory with varying $G$ the
equations for the comprehensive variational principle $\delta \widetilde{I}%
_{tot}=0$.

The variational equation for $\phi $ remains unchanged as (3.17) because of
(10.10). The variational equations for $g_{\mu \nu }$\ are
\begin{equation*}
N_{4}^{4}=-8\pi (\widetilde{\rho }_{m}+\widetilde{\rho }_{r}),\text{ \ }%
N_{1}^{1}=8\pi \widetilde{p}_{r}=8\pi \widetilde{\rho }_{r}/3\text{ \ (11.1)}
\end{equation*}%
We give these equations in full, cf. (4.8), (4.9) and (4.10)
\begin{eqnarray*}
\Phi &=&6[\overset{\cdot \cdot }{R}/R+\overset{\cdot }{R}^{2}/R^{2}]\phi \\
\text{ }+8[\overset{\cdot \cdot }{\phi }+3(\overset{\cdot }{R}/R)\overset{%
\cdot }{\phi }] &=&0\ \ \ \ \ \ \ \ \ \text{\ \ }\ \text{(11.2)}\ \ \ \ \ \
\ \
\end{eqnarray*}%
\begin{eqnarray*}
N_{1}^{1} &=&\phi ^{2}[-\overset{\cdot }{R}^{2}/R^{2}-2\overset{\cdot \cdot }%
{R}/R]-(\phi ^{2})^{\cdot \cdot }-2(\phi ^{2})^{\cdot }\overset{\cdot }{R}/R
\\
+4\overset{\cdot }{\phi }^{2} &=&(8\pi /3)\widetilde{\rho }%
_{r}(t_{0})R^{4}(t_{0})/R^{4}\text{ \ \ \ \ \ \ \ \ \ \ (11.3)}
\end{eqnarray*}%
\begin{eqnarray*}
N_{4}^{4} &=&\phi ^{2}[-3\overset{\cdot }{R}^{2}/R^{2}]-3(\phi ^{2})^{\cdot }%
\overset{\cdot }{R}/R-4\overset{\cdot }{\phi }^{2} \\
&=&-8\pi \lbrack \widetilde{\rho }_{m}(t_{0})R^{3}(t_{0})/R^{3}+\widetilde{%
\rho }_{r}(t_{0})R^{4}(t_{0})/R^{4}]\text{ \ \ \ \ \ (11.4)}
\end{eqnarray*}%
The differential identity, cf. (3.6), $N_{4;\beta }^{\beta }+\Phi
\phi _{;4}=0$ shows that among the three equations (11.2,3,4) only
two are independent, the condition for compatiability on the right
hand side being separately guaranteed for matter and for
electromagnetic radiation. The combination $N_{\nu }^{\nu }+\Phi
\phi $ cf (3.18) remains unchanged and can be integrated to give,
cf. (4.2),
\begin{equation*}
(\phi ^{2})^{\cdot }R^{3}/R^{3}(t_{0})=-8\pi \widetilde{\rho }_{m}(t_{0})t%
\text{ \ \ (11.5)}
\end{equation*}%
which can be used in place of any equation among (11.2,3,4). We may take as
the two independent equations (11.4) and (11.5), and use the new variables
\begin{equation*}
\tau =t/t_{0},\xi =R/R(t_{0}),\eta =\phi ^{2}/\phi ^{2}(t_{0})\text{ (11.6)}
\end{equation*}%
\ \ \ \ \ \ \ \ and parameters
\begin{eqnarray*}
8\pi t_{0}^{2}\widetilde{\rho }_{m}(t_{0})/\phi ^{2}(t_{0}) &=&8\pi
t_{0}^{2}G(t_{0})\rho _{m}(t_{0})=A\text{ \ \ \ \ \ \ \ \ \ \ \ (11.7) \ } \\
\widetilde{\rho }_{r}(t_{0})/\widetilde{\rho }_{m}(t_{0}) &=&\rho
_{r}(t_{0})/\rho _{m}(t_{0})=\varepsilon _{r/m}\sim (10^{-2}\text{ to }%
10^{-4})\text{\ \ \ (11.8)}
\end{eqnarray*}%
Then (11.4) and (11.5) become, with circle denoting $d/d\tau $,
\begin{equation*}
-3\eta \overset{\circ }{\xi }^{2}/\xi ^{2}-3\overset{\circ }{\eta }\overset{%
\circ }{\xi }/\xi -\overset{\circ }{\eta }^{2}/\eta =-A/\xi ^{3}-\varepsilon
_{r/m}A/\xi ^{4}\text{ (11.9)}
\end{equation*}%
\begin{equation*}
\overset{\circ }{\eta }\xi ^{3}=-A\tau \text{ \ (11.10)}
\end{equation*}%
The solution for matter dominated universe given in \S 4 is for $\varepsilon
_{r/m}=0$ that $\xi =\tau $, $\eta =1/\tau $, and $A=1.$\ This solution
satisfies not only the ``initial" conditions at $\tau =1$ that $\xi =1$ and $%
\eta =1$ but also the ``final" condition at $\tau =0$ that $\xi
=0$. The extra condition serves to determine the parameter $A$. We
shall leave the mathematical problem of solving these equations
(11.9), (11.10) but only note that the asymptotic solution of
these equations for large $\xi $\ is that of matter dominated
universe. Only with the help of the asymptotic solution which is
simple enough to reveal the cosmical combined conservation laws
that we find the way to modify or to generalize Einstein's theory
of general relativity to work to cosmological time. For small
values of $\tau ,$ at $\tau \sim \varepsilon _{r/m}$, the
deviation from the asymptotic solution will be appreciable. For
smaller values of $\tau $ refinement of
the action integral for matter with corresponding refinement of $\widetilde{T%
}_{\mu \nu }$ need to be considered to take account of the internal energy
and pressure, which at thermal equilibrium can be treated according to the
principle of statistical mechanics.

\section{Statistical mechanics}

The fundamental natural constant in statistical mechanics is the Boltzmann
constant $k_{B}$, in this section we shall see how it evolves in
cosmological long time.\ For this purpose we need to consider for example
black body radiation where $k_{B}$ appears in Planck's formula and in the
energy density of black body radiation.
\begin{equation*}
\rho _{r}=\int_{0}^{\infty }\frac{8\pi h\nu ^{3}/c^{3}}{\exp (h\nu /k_{B}T)-1%
}d\nu =\frac{\pi ^{2}}{15c^{3}}\frac{k_{B}^{4}}{\hslash ^{3}}T^{4}\text{ \ \
(12.1)}
\end{equation*}%
Comparing this with the cosmic combined expression
\begin{equation*}
\widetilde{\rho }_{r}=\int_{0}^{\infty }\frac{8\pi \widetilde{h}\nu
^{3}/c^{3}}{\exp (\widetilde{h}\nu /\widetilde{k}_{B}T)-1}d\nu =\frac{\pi
^{2}}{15c^{3}}\frac{(\widetilde{k}_{B})^{4}}{(\widetilde{\hslash })^{3}}T^{4}%
\text{ \ (12.2)}
\end{equation*}%
and noting that like $\rho _{m}$ (cf. (11.1) and (2.5)) we must have \ $%
\widetilde{\rho }_{r}=G^{1+1/n}\rho _{r}$, as we know that electromagnetic
radiation can be regarded as matter, namely photons. From this we obtain
from (12.1) (12.2) and (8.7) the similar relation%
\begin{equation*}
\widetilde{k}_{B}=G^{1+1/n}k_{B}\text{ \ \ (12.3)}
\end{equation*}%
Thus we have the general rule for c.c. energy, cf. (2.5), (8.7), (12.3), and
also cf.(8.8) for the c.c. Coulomb energy,
\begin{equation*}
\widetilde{E}=G^{1+1/n}E,\text{ (}\widetilde{E}=\widetilde{m}c^{2},%
\widetilde{h}\nu ,\widetilde{k}_{B}T,\widetilde{e}\widetilde{\kappa }_{4}%
\text{) \ (12.4)}
\end{equation*}%
where $c^{2},\nu ,T$ do not change. Since energy, momentum and
stress form a four dimensional tensor, the above relation (12.4)
holds too for the momentum
$\widetilde{p}_{x},\widetilde{p}_{y,}\widetilde{p}_{z}$\ and\ the
pressure $\widetilde{P}$, while velocity $u_{x},u_{y},u_{z}$ and
volume $V$
do not change. This makes the Boltzmann factor%
\begin{equation*}
\exp (-\widetilde{E}/\widetilde{k}_{B}T)=\exp (-E/k_{B}T)\text{ \ (12.5)}
\end{equation*}
the same and the counting of phase cells
\begin{equation*}
Vd\widetilde{p}_{x}d\widetilde{p}_{y}d\widetilde{p}_{z}/(\widetilde{h}%
)^{3}=Vdp_{x}dp_{y}dp_{z}/h^{3}\text{ (12.6)}
\end{equation*}
also the same, so for canonical ensemble of systems at local thermal
equilibrium, the partition function (Zustandsumme) $\widetilde{Z}=Z$ and
hence the distribution function is the same with or without the cosmic
combination factor. For example, for Bose-Einstein statistics, the
distribution function of photons is given by, cf (12.1) and (12.2), where
the factor two accounts for the two polarizations
\begin{equation*}
\frac{(2)4\pi \nu ^{2}/c^{3}}{\exp (\widetilde{h}\nu /\widetilde{k}_{B}T)-1}=%
\frac{(2)4\pi \nu ^{2}/c^{3}}{\exp (h\nu /k_{B}T)-1}\text{ \ (12.7)}
\end{equation*}%
\

We shall not go into the details of the early universe but suffice it to
note that at high temperature the internal energy and pressure must be taken
into account. In Einstein's theory the action integral $I_{m}$ (E2.3a) for
matter without internal stress is replaced by that of perfect fluid $I_{fl}$%
, \{we follow here Fock's treatment in his book \cite{Fock} space
time and gravitation, eq.(48.28) et seq.\}, the cosmic combined
expression in our theory is given accordingly by
\begin{eqnarray*}
\widetilde{I}_{fl} &=&-\int F(\widetilde{\rho }_{m})\surd (-g)d^{4}x \\
&=&-\int \widetilde{\rho }_{m}(1+\widetilde{\epsilon }_{m})\surd (-g)d^{4}x%
\text{ (12.8)}
\end{eqnarray*}%
Here $\widetilde{\epsilon }_{m}$ denotes the c.c. internal energy per unit
c.c. mass $\widetilde{\rho }_{m}$ of the ideal fluid and is considered as a
function of\ the latter only, as we are dealing with adiabatic processes in
which the c.c. entropy per unit c.c. mass, denoted here by $\widetilde{s}_{m}
$ is kept constant, i.e.
\begin{equation*}
d\widetilde{\epsilon }_{m}+\widetilde{P}d(1/\widetilde{\rho }_{m})=Td%
\widetilde{s}_{m}=0\text{ \ (12.9)}
\end{equation*}%
To find the variation of $\widetilde{I}_{fl}$ we express $\widetilde{\rho }%
_{m}$\ in terms of $\widetilde{\wp}^{\mu }$\ by (2.4a) and obtain
\begin{eqnarray*}
\delta \widetilde{I}_{fl} &=&-\int \frac{dF}{d\widetilde{\rho }_{m}}\delta (%
\widetilde{\rho }_{m}\sqrt{-g})d^{4}x \\
&&+\int (\widetilde{\rho }_{m}\frac{dF}{d\widetilde{\rho }_{m}}-F)(\delta
\sqrt{-g})d^{4}x \\
&=&-(1/2)\int (\widetilde{T}_{fl}^{\mu \nu }\delta g_{\mu \nu })\surd
(-g)d^{4}x \\
&&-\int \frac{dF}{d\widetilde{\rho }_{m}}u_{\mu }\delta \widetilde{\text{$%
\wp $}}^{\mu }d^{4}x\text{ \ (12.10)}
\end{eqnarray*}%
with
\begin{eqnarray*}
\widetilde{T}_{fl}^{\mu \nu } &=&\widetilde{\rho }_{m}\frac{dF}{d\widetilde{%
\rho }_{m}}u^{\mu }u^{\nu }-(\widetilde{\rho }_{m}\frac{dF}{d\widetilde{\rho
}_{m}}-F)g^{\mu \nu } \\
&=&[\widetilde{\rho }_{m}(1+\widetilde{\epsilon }_{m})+\widetilde{P}]u^{\mu
}u^{\nu }-\widetilde{P}g^{\mu \nu }\text{ (12.11)}
\end{eqnarray*}%
Using (3.11) in compliance with the modified constraint (2.3b) we obtain
from the second part of (12.10) the following equations of motion for the
ideal fluid
\begin{eqnarray*}
&&\widetilde{\rho }_{m}\frac{dF}{d\widetilde{\rho }_{m}}u^{\nu }u_{\mu ;\nu
}+\widetilde{\rho }_{m}u^{\nu }u_{\mu }(\frac{dF}{d\widetilde{\rho }_{m}}%
)_{;\nu }-\widetilde{\rho }_{m}(\frac{dF}{d\widetilde{\rho }_{m}})_{;\mu }%
\text{ } \\
&=&[\widetilde{\rho }_{m}(1+\widetilde{\epsilon }_{m})+\widetilde{P}]u^{\nu
}u_{\mu ;\nu }+u^{\nu }u_{\mu }\widetilde{P}_{;\nu }-\widetilde{P}_{;\mu }=0%
\text{ \ (12.12)}
\end{eqnarray*}%
One can verify the covariant divergence identity
\begin{eqnarray*}
\widetilde{T}_{fl.\mu ;\nu }^{\nu } &=&(\widetilde{\rho }_{m}u^{\nu })_{;\nu
}\frac{dF}{d\widetilde{\rho }_{m}}u_{\mu }+\widetilde{\rho }_{m}u^{\nu }(%
\frac{dF}{d\widetilde{\rho }_{m}}u_{\mu })_{;\nu }
-(\widetilde{\rho }_{m}\frac{dF}{d\widetilde{\rho }_{m}}-F)_{;\mu } =0%
\text{ \ \ \ \ \ \ \ \ \ \ \ \ \ (12.13)}
\end{eqnarray*}%
with the help of (12.12) and (2.3b) (2.4b). We note that $\widetilde{I}_{fl}$%
, like $\widetilde{I}_{m}$, does not involve $\phi $, hence
contributes nothing to the variational equation with respect to
$\phi $.

In the specific internal energy $\widetilde{\epsilon }_{m}$,\ being energy
per unit mass, according to (12.4), the cosmic combination factor will
cancel. Thus we expect $\widetilde{\epsilon }_{m}=\epsilon _{m}$. This can
be easily verified, e.g. for an ideal gas at low temperature. With Maxwell's
law for the distribution of velocity $\exp [-\widetilde{m}%
(u_{x}^{2}+u_{y}^{2}+u_{z}^{2})/(2\widetilde{k}_{B}T)]du_{x}du_{y}du_{z}$
in which $\widetilde{m}$\ and\ $\widetilde{k}_{B\text{ }}$can be
simultaneously
be replced by $m$ and $k_{B}$, the translational internal energy is $%
\widetilde{\rho }_{m}\widetilde{\epsilon
}_{m}=3n_{m}\widetilde{k}_{B}T/2$\ while $\widetilde{\rho
}_{m}=n_{m}\widetilde{m}$\ where $\widetilde{m}$ being the c.c.
conservered mass of one molecule, and $n_{m}\ $the\ number density
of the molecules. Similarly $\widetilde{\epsilon }_{m}=\epsilon
_{m}$ will hold for relativistic gas at high temperatures.

\section{Summary and remark}

From our theory with varying $G$\ we have sysmatically found the cosmic
combined natural constants $\widetilde{m},\widetilde{\hslash },\widetilde{k}%
_{B},\widetilde{e}$
\begin{equation*}
\widetilde{m}/m=\widetilde{\hslash }/\hslash =\widetilde{k}%
_{B}/k_{B}=[G/G_{0}]^{1+1/n},\widetilde{e}/e=[G/G_{0}]^{1/2+1/n}\text{ \
(13.1)}
\end{equation*}%
which remain to be constant in the long. Our present fundamental principles
of quantum mechanics and statistical mechanics work in the long with $%
\widetilde{m}$, $\widetilde{\hslash }$\ and $\widetilde{k}_{B}$.\ It is due
to the old age of the present universe that the combination factors $%
G^{1+1/n}$ involved in these c.c. constants vary very slowly by now that we
take $m$, $\hslash $ and $k_{B}$\ as natural constants. The law of
electromagnetism in the long differs from the law at present by a dielectric
and magnetic permittivity $\varepsilon =\mu ^{-1}=$ $[G/G_{0}]^{1/n}$.\ The
velocity of light $c$, the Compton wave length for a particle $\widetilde{%
\hslash }/\widetilde{m}c$,\ the fine structure constant\ $\widetilde{e}%
^{2}/(\varepsilon \widetilde{\hslash }c)$ and the proton-electron mass ratio
$\widetilde{m}_{p}/\widetilde{m}_{e}$ all remain unchanged during the
evolution.

For phenomena occurring outside the gravitating body like the
crucial tests our theory to the non-cosmological approximation
reduces to Einstein's theory. Difference arises for phenomena
occurring inside the gravitating body, including the case of
cosmology. E.g. for matter-dominated universe our theory with
varying $G$\ yields for the expansion law $R\varpropto t$ between
the radius and the age of universe so as to be in agreement with
the empirical large number equality $GM\varpropto t$, while in
Einstein's theory with constant $G$ and conservation of mass we
have $R\varpropto t^{2/3}$ with $GM=\textrm{constant}$\ of course.
In our theory with varying $G$\ , a tensor term automatically
arises from the spatial and temporal derivatives of $G$. This
tensor automatically vanishes in a region where $G$ is constant,
as is the case with the exterior solution of the sun when we
consider the crucial tests of general relativity. In the problem
of non-static homogeneous cosmology this tensor with non-vanishing
components $\Lambda _{1}^{1}=\Lambda _{2}^{2}=\Lambda
_{3}^{3}=1/t^{2},$ $\Lambda _{4}^{4}=2/t^{2}$ (Or less probably
$\Lambda _{1}^{1}=\Lambda _{2}^{2}=\Lambda _{3}^{3}=2/t^{2},$
$\Lambda _{4}^{4}=1/t^{2}$) takes the place of cosmological
constant. It seems desirable to make an analysis of the
observational data from the beginning according to the present
theory with varying $G$.\ It would be also interesting to apply
the theory with varying $G$\ to the interior solution of some high
density objects where spatial variation of $G$\ is appreciable.

\section{ACKNOWLEDGMENTS}

I wish to thank Tu Zhan-chun for calling my attention to a recent
review article on the fundamental constants and their variation by
Jean-Philippe Uzan \cite{Uzan}. No apparent discrepancy is found
between the results summarized there and those given by the
formulae of the present theory. I also thank him for his help in
making this manuscript as latex file.

\end{document}